\documentclass[12pt]{revtex4}
\usepackage{color}
\usepackage{amsmath,amssymb,amsfonts,dcolumn,color,graphicx,graphics,latexsym,epsfig}
\def\beq{\begin{equation}}
\def\eeq{\end{equation}}
\def\barr{\begin{array}}
\def\earr{\end{array}}
\def\dis{\displaystyle}

\begin{document}

\title{Radion stability and induced, on-brane geometries in an 
effective scalar-tensor theory of gravity}

\author{Sayan Kar${}^\dagger$, Sayantani Lahiri${}^{\#}$ and Soumitra SenGupta}
\email{sayan@iitkgp.ac.in,sayantani.lahiri@gmail.com,tpssg@iacs.res.in}
\affiliation{${}^\dagger$Department of Physics {\it and} Center for Theoretical Studies \\Indian Institute of Technology, Kharagpur, 721 302, India.}
\affiliation{${}^{\#}$Relativity and Cosmology Centre, Department of Physics,
 Jadavpur University, Raja Subodh Chandra Mullick Road, Jadavpur, Kolkata- 700 032, India}
\affiliation{${}^{*}$Department of Theoretical Physics, Indian Association for the
Cultivation of Science
\\ 2A and 2B Raja S.C. Mallick Road, Jadavpur, Kolkata 700 032, India.}

\begin{abstract}
\noindent About a decade ago, using a specific expansion scheme, effective, 
on-brane 
scalar tensor theories of gravity were proposed by Kanno and Soda
( Phys.Rev. {\bf D 66} 083506 ,(2002)) in the
context of the warped two brane model of Randall--Sundrum.
The inter-related effective theories on both the branes were derived with the
space-time dependent radion field playing a crucial role. Taking a re-look 
at this effective theory,
we find cosmological and spherically symmetric, static solutions sourced by
a radion--induced, effective stress energy, as well as additional, on-brane
matter.
The distance between the branes (governed by the time or
space dependent radion)
is shown to be stable and asymptotically non-zero, thereby setting aside
any possibility of brane collisions. It turns out that
the inclusion of on-brane matter plays a decisive role in stabilising the radion - a fact which we demonstrate 
through our solutions.
\end{abstract}


\maketitle

\newpage

\section{Introduction}
\noindent The possible existence of extra spatial dimensions is now a 
well-known 
theoretical assumption where our four dimensional world is considered to be 
a $3$-brane embedded in a higher dimensional spacetime. Such a description 
emerges naturally in the backdrop of various string-inspired models \cite{string}. 
Moreover, extra dimensional models were developed as a non-supersymmetric,
alternative approach in tackling the well-known 
fine tuning/gauge hierarchy problem in the regime of the 
Standard Model of particle physics. 
It became more and more  evident that gravity may become an integral part 
to address issues on physics beyond the Standard model.\\ 
The extra dimensional models can broadly be classified into those having 
large compact radii \cite{arkani} or having small 
compact radii \cite{randall}. Regarding their geometry, these models are 
generally compactified under various topological setups. 
The uncompactified, four dimensional spacetime then emerges as a low energy effective theory which contains signatures of the 
higher dimensional theory.\\
However, among all models proposed so far, we will confine ourselves to 
the Randall-Sundrum (RS) model \cite{randall} which has two $3$-branes, with 
equal 
and opposite brane tensions, embedded in a five dimensional spacetime.
This model was initially developed to combat the unnatural fine tuning involved in determining the mass of the Higgs boson. 
While determining the theoretically predicted mass of the Higgs boson ($100-125$ GeV) from higher order self energy calculations, this boson gets quantum corrections typically 
of the order of the Planck energy scale. 
As a result, an extreme fine tuning needs to be carried out at every 
order of perturbation theory to obtain the theoretically predicted value.
This fine tuning is often known as the Higgs mass hierarchy problem or 
naturalness problem in particle physics.
Without introducing any intermediate scale in the theory, the RS model 
successfully resolved the fine tuning problem by exponentially suppressing 
all mass scales on one of the $3$-branes, known as the visible brane. 
Thus the entire low energy theory is reproduced on the  negative tension 
visible brane at TeV scale. 
By far, this is one of the most successful approach for addressing the naturalness 
problem for a constant inter-brane separation. \\
However the RS model suffered from the stabilization problem. 
In the absence of any stabilization scheme, the two brane system can collapse 
under the influence of equal and opposite brane tensions. 
Therefore, a reasonably generic method for stabilising the brane 
separation distance $r_c$ or the modulus field, was proposed by 
Goldberger and Wise \cite{GW}
in which a stabilizing potential for the modulus field is generated by a 
$5D$ bulk scalar field with appropriate value at the boundary. 
The minimum of the modulus potential corresponds to 
the vev of the modulus field ($ k r_c$). From this condition the vev of the 
modulus field can be set as $kr_c \simeq 11.5$ ( to resolve the naturalness problem) without any fine-tuning of the $4D$ parameters. 
In other words, the stabilisation is achieved without sacrificing the conditions necessary to solve  the gauge hierarchy problem. \\
Besides offering explanations to the problems beyond the 
Standard Model of particle physics, the RS model has attracted the 
attention of 
cosmologists due to its unique interpretation of the 
cosmological constant fine tuning problem.
Therefore, over the last decade, various cosmological and astrophysical 
issues like galaxy formation, existence of anisotropies in cosmic microwave 
background, dark energy and dark matter, black hole formation 
have been extensively studied in the context of the RS two-brane model (see 
\cite{maartens} and references therein).\\
In the present paper, we consider the effective, on--brane, scalar-tensor 
theories formulated by Kanno and Soda \cite{kanno} where the radion field, 
which measures the inter-brane separation between the visible brane and the 
Planck brane is not a constant quantity. In fact, while studying the 
cosmological solution on the visible or the Planck brane, the 
radion is taken as a time dependent field. Similarly, for spherically symmetric, static
on-brane geometries, the radion field depends on the radial
coordinate. The spatial or temporal dependence of the radion therefore
leads to the requirement that it must be non-zero everywhere in order to
avoid brane collisions. We are able to demonstrate that by assuming the existence of 
on-brane matter, a stable non-zero distance between the branes is
possible. 

\noindent In the next section we provide an overview of the effective
scalar-tensor theories proposed by Kanno and Soda \cite{kanno}. 
Subsequently in Section III, we deal with cosmological solutions and
in Section IV we look at spherically symmetric solutions. In the
last Section, we provide our summary and conclusions.

\section{Gradient expansion scheme and the Kanno-Soda effective theory}

\noindent Let us now briefly discuss the low energy effective theory on a 
$3$-brane 
developed by Kanno and Soda \cite{kanno} in the context of the two-brane model 
developed by Randall and Sundrum. The two $3$-branes being $Z_2$ symmetric 
are located at orbifold fixed points $y=0$ and $y=l$ such that the geometry 
under consideration in this model is:  $M^{1,3}\times S^1/Z_2$. Our Universe 
is assumed to be on the visible $3$-brane which is a hypersurface 
embedded in a five dimensional AdS bulk filled with only a 
$5D$ bulk cosmological 
constant. The bulk curvature scale is $l$. Typically, in the RS model, the 
Einstein equations are determined by keeping the inter-brane distance fixed 
and considering a flat $3$-brane. However, the scenario drastically changes 
once the inter-brane separation distance or the proper length becomes a 
function of the spacetime co-ordinates and the on-brane geometry is curved.  
These generalizations are incorporated while 
deriving the effective equations of motion on a $3$-brane \cite{kanno}. 
Beginning with \cite{sms} there has been a lot of work on 
the effective Einstein equations on the brane under various assumptions
\cite{eee}.
In fact, the effective equations for the two-brane system as obtained in
\cite{kanno} has also been re-derived in a different  approach in \cite{tskk}. 
An interesting recent work on slanted warped extra dimensions and its
phenomenological consequences appeared in \cite{slant}. 

\noindent In order to determine the effective theory, we assume the following 
five dimensional action and a five dimensional metric with a spacetime varying 
proper distance between the two $3$ branes. The action functional is given as
\beq
S\,=\,\frac{1}{2 \kappa^2}\int d^5 x \sqrt{-g}\left( \textit{R} +\dis\frac{12}{l^2}\right)-\dis\Sigma_{i = a,b}\,\sigma_{\textit{i}} \int d^4 x \sqrt{-g^{\textit{i}\,brane}}\,+\,\dis\Sigma_{i = a,b} \int d^4 x \sqrt{-g^{\textit{i}\,brane}}\,L^{\textit{i}}\,_{matter}
\eeq
where the tensions on the Planck brane and visible brane are respectively given by  $\sigma_a = \dis\frac{6}{\kappa^2 l}$ and $\sigma_b = - \dis \frac{6}{\kappa^2 l}$. Let us consider the most general $5$D line element,
\beq
ds^2\,=\,e^{2 \phi(x)}\, dy^2 \,+\,g_{\mu \nu}(y, x^{\mu})\,dx^{\mu} dx^{\nu}
\eeq
where $\kappa^2$ is five dimensional gravitational coupling constant.
Since both cosmological and astrophysical solutions that we consider in the present case occur at energy scales 
much lower than that of the Planck scale, therefore in the effective theory approach
the brane curvature radius $L$ is much large compared to bulk curvature $l$. As a result, perturbation theory can be used with a dimensionless perturbation parameter $\epsilon$ such that $\epsilon = (\frac{l}{L})^2 << 1$.
This method, called the gradient approximation scheme, is a metric-based 
iterative method in which the bulk metric and extrinsic curvature are expanded with increasing order of $\epsilon$ in perturbation theory. The effective 
Einstein equations on a brane are determined with the solutions of these 
quantities and the junction conditions. In this method, the RS fine 
tuning condition is reproduced at the zeroth order when the inter-brane 
separation is constant and the two $3$-branes are characterised by opposite 
brane tensions.
The effective Einstein equations are then obtained at the first order
incorporating non-zero contributions of the radion field and brane matter. 
Using the gradient expansion scheme, the effective Einstein equations on the 
visible brane are as follows: \cite{kanno}
\begin{eqnarray}
G_{\mu \nu} = \frac{\kappa^2}{l\Phi} T^{b}_{\mu \nu} + \frac{\kappa^2\,(1 + \Phi)}{l\Phi} 
T^{a}_{\mu \nu} + \frac{1}{\Phi}\left (\widetilde{\nabla}_{\mu} \widetilde{\nabla}_{\nu} \Phi - f_{\mu \nu} \widetilde{\nabla}^{\alpha} \widetilde{\nabla}_{\alpha} \Phi\right ) \nonumber \\  -\frac{3}{2\Phi(1+\Phi)} \left (\widetilde{\nabla}_{\mu}\Phi\widetilde{\nabla}_{\nu}\Phi
-\frac{1}{2} f_{\mu \nu} \widetilde{\nabla}^{\alpha}\Phi\widetilde{\nabla}_{\alpha} \Phi\right )  \label{EE}
\end{eqnarray}
where $\Phi = e^{2\frac{d}{l}}-1 $ and $d$ is the proper distance between the branes which in general is a spacetime dependent quantity. $\kappa ^2$ is the $5D$ gravitational coupling constant. $T^{a}_{\mu \nu}$, $T^{b}_{\mu \nu}$ are the matter on the Planck brane and the visible brane respectively. All covariant derivatives in the above expression are defined w.r.t. the metric
on the visible brane (denoted by the superscript `b') given by $f_{\mu\nu}$. \\
The proper distance, a spacetime dependent function, between the two $3$-branes in the interval $y=0$ and $y=l$ is defined as :
\beq
d(x)\,=\, \int^{l}_{0}e^{\phi (x)} dy
\eeq
and the corresponding equation of motion of the scalar field on the negative tension brane is given by,
\begin{equation}
\widetilde{\nabla}^{\alpha}\widetilde{\nabla}_{\alpha}\Phi =\frac{\kappa^2}{l}\frac{T^a + T^b}{2\omega+3} - 
\frac{1}{2\omega +3} \frac{d\omega}{d\Phi} (\widetilde{\nabla}^{\alpha}\Phi)(
\widetilde{\nabla}_{\alpha}\Phi )
\end{equation}
Here $T^{a}$ and $T^{b}$ are traces of energy momentum tensors on Planck brane and visible brane respectively. The coupling function $\omega({\Phi})$ in terms of $\Phi$ can be expressed as,
\begin{equation}
\omega (\Phi) = -\frac{3\Phi}{2(1+\Phi)}
\end{equation}

\noindent It is however known that the gravity on both the branes are not independent. The dynamics on the Planck brane situated at $y=0$ is related to that of the visible brane by the following transformation \cite{kanno} :
\beq
\Phi(x)\,=\,\frac{\Psi}{1 - \Psi }
\eeq
where $\Psi$ is the radion field defined on Planck brane. Now, the induced metric on the visible brane can be expressed in terms of $\Psi$ as,
\beq
g^{b-brane}_{\mu \nu}\,=\,(1 - \Psi)\,[h_{\mu \nu} + g^{(1)}_{\mu \nu}(h_{\mu \nu},\Psi, T^{a}_{\mu \nu},T^{b}_{\mu \nu},y=l)]
\eeq
where $g^{(1)}_{\mu \nu}$ is the first order correction term. \\
It is to be noted that in the subsequent calculations we will assume that the on-brane stress
energy is present only on the `b' brane i.e. on the visible brane.

\noindent An important feature of the effective equations given above is
that unlike the ones derived in \cite{sms} there is no non-local contribution
(bulk-Weyl-dependent ${\cal E}_{\mu\nu}$ \cite{sms}) from bulk geometry.

\section{Cosmological solutions}
\noindent In order to study the cosmological solution on the negative tension,
visible brane, we assume the radion field to be time dependent. 
Therefore, the 
proper distance between the orbifold fixed points i,e. $y=0$ to $y=l$ is given by,
\beq
d(t)\,=\,\int_{y=0}^{y=l}e^{\phi(t)}dy \,=\, l e^{\phi(t)}
\eeq 
The Friedmann-Robertson-Walker (FRW) solutions of the Einstein equations 
can be obtained for
three different types of spatial curvature, $k = -1,0,1$. In this 
section, we study the solutions corresponding to each of these values of $k$ separately.
The FRW metric with a non-zero spatial curvature is given by : 
\beq
ds^2\,=\,-dt^2\,+\,a^2(t)\left[\frac{dr^2}{1-k\,r^2}+ r^2(d \theta ^2 + \sin \theta ^2 d \phi ^2) \right]    \label{lin_ele}
\eeq
where $r$, $\theta$, $\phi$ are the radial co-ordinates and $a(t)$ is the scale factor to be determined. Substituting the above metric in eqn.{(\ref{EE})}, the Einstein's equations with spatial curvature $k$ are obtained as follows :
\beq
3\,\,\left (\frac{\dot a}{a}\right )^2 + 3 \,\,\frac{\dot a}{a} \frac{\dot \Phi}{\Phi} + \frac{3k}{a^2}
+ \frac{{3\dot\Phi}^2}{4\Phi(1+\Phi)} = \frac{\kappa^2}{l\Phi} \,\, \rho  \label{20}
\eeq
\beq
2\frac{\ddot a}{a} +\left (\frac{\dot a}{a}\right )^2 + \frac{k}{a^2} - \frac{\dot a}{a} \frac{\dot \Phi}{\Phi}
-\frac{{\dot\Phi}^2}{4\Phi(1+\Phi)} = \frac{\kappa^2}{3 l}(-\rho + 3p) - \frac{\kappa^2}{3 l\Phi} \rho \label{9}
\eeq
and the scalar field equation is given by :
\begin{equation}
\ddot \Phi + 3\frac{\dot a}{a}\dot\Phi = \frac{\kappa^2(\rho - 3p)(1+\Phi)}{3 l} + \frac{{\dot\Phi}^2}{2(1+\Phi)} \label{8}
\end{equation}
where an overdot represents derivative with respect to time $t$. 
It is to be noted that eqn.{(\ref{9})} is obtained by substituting {(\ref{8})}  in $(ii)$ component of the Einstein's equations. The scalar field equation is found to be independent of spatial curvature $k$ and hence the equation remains same for any value of $k$. However, the scalar field profile is different for different $k$ values due to the different functional forms of $a(t)$. \\
Let us now consider each value of $k$ separately and study the cosmological 
solution in the presence of a radion field with a time dependence. 
\subsection{Spatially flat solution ($ k = 0 $)}
\noindent To construct a spatially flat FRW Universe on the visible brane 
in the presence of a time dependent radion field, we consider the line element 
given by eqn.{(\ref{lin_ele})}, which, for $k = 0 $ reduces to:
\beq
ds^2\,=\,-dt^2 + a^2(t)\,\delta_{ij}\,dx^{i}dx^{j}
\eeq
We initially assume that both the $3$-branes are devoid of brane energy 
densities and pressures. Therefore when $\rho =0 = p$, eqn.{(\ref{8})} can be re-expressed in terms of first integral of the $\Phi$ equation. The  
scalar field equation reduces to:
\beq
\dot\Phi^2 =\frac{C_1^2}{a^6} \left (1+\Phi\right )    \label{3}
\eeq
After substituting $k=0$ and $\rho =0 = p$ in eqn.{(\ref{20})}, eqn.{(\ref{9})} and adding the two equations we get,
\beq
\frac{\ddot a}{a} +\left (\frac{\dot a}{a}\right )^2=0  \label{4}
\eeq
Integrating eqn.{(\ref{4})}  we get, 
\beq
a(t) = \sqrt{2 \tilde{C}_1 t - C_2}
\eeq
Now, we can choose the dimensionful factor $\tilde{C}_1 =1$ by a 
scaling choice, so that the solution of scale factor is re-written as,
\begin{eqnarray}
a(t) = \left (2t-C_2\right )^{\frac{1}{2}}   \label{5}
\end{eqnarray}
where $C_2$ is a constant of integration. 
Substituting eqn.{(\ref{3})} and the scale factor into eqn.{(\ref{20})} (with $k=0$) and then integrating it gives the solution for time dependent scalar field as:
\begin{eqnarray}
\Phi (t) = \frac{C_1^2}{4(2t-C_2)} + \frac{C_1}{(2t-C_2)^{\frac{1}{2}}}  \label{7}
\end{eqnarray}
where $C_1$ is a non-zero constant with dimensions of $L^{\frac{1}{2}}$. 
The constant $C_2$ may be set to zero by time translation so that
$a(0)=0$. However, $C_1$ must be strictly non-zero so that the scalar field $\Phi(t)$ remains non-zero as well. From the above solution of $\Phi(t)$ we can construct the proper distance $d(t)$ as given below: 
\begin{equation}
d(t) = \frac{l}{2} \ln \left[ 1+\frac{C_1^2}{8t} + \frac{C_1}{\sqrt{2t}}\right ]
\end{equation}
The above solution indicates that the scale factor has a decelerating (but expanding) nature and the scalar field approaches zero in the later time 
whereas it is large in the early universe. The obtained solution is similar to that of the FRW radiation--dominated universe. However, $d(t)$, which measures the inter-brane distance, tends to
zero in the limit $t\rightarrow \infty$ thereby indicating an instability. \\
Let us now consider a perfect fluid but with the equation of state $p=\frac{\rho}{3}$ and then construct the solutions. 
The traceless property of the energy momentum tensor for a perfect fluid with $p=\frac{\rho}{3}$ offers some simplifications.  With the above mentioned equation of state, addition of eqn.{(\ref{20})} and eqn.{(\ref{9})} for $k=0$ produces same differential equation for the scale factor $a(t)$ as before and hence the same solution which is:
\beq
a(t)=\sqrt{2t}
\eeq
where we have set the constant $C_2=0$. Using the scale factor derived above in eqn.{(\ref{3})}, the solution of the scalar field can now be written as,
\begin{equation}
\Phi(t) = \frac{C_1^2}{8t} \pm \frac{C_1 A}{2\sqrt{2t}}+\frac{A^2-4}{4} \label{6}
\end{equation}
where we now have an extra parameter $A$. 
Now using the solutions of $a(t)$ and $\Phi(t)$, the energy density on the visible brane is given by,
\begin{equation}
\rho (t) =  \frac{l}{\kappa^2}\,\,\frac{3(A^2-4)}{16 t^2}
\end{equation}
We note that when $A=2$, eqn.{(\ref{6})} exactly reduces to the solution of $\Phi(t)$ given in eqn.{(\ref{7})} (with $C_2 = 0$) which is the scalar field solution in the absence of the brane matter on both the $3$-branes. The nature of the variation of  $\Phi(t)$ versus $t$ is shown in Figure 1 where $A = 2$, $C_1=2\sqrt{2}$ (red curve) and $A=3, C_1=2\sqrt{2}$ (green curve). The horizontal line 
(blue) shows the non-zero asymptotic value of $\Phi(t)$ when brane-matter
is present.
\begin{figure}[h]
\includegraphics[width=28pc]{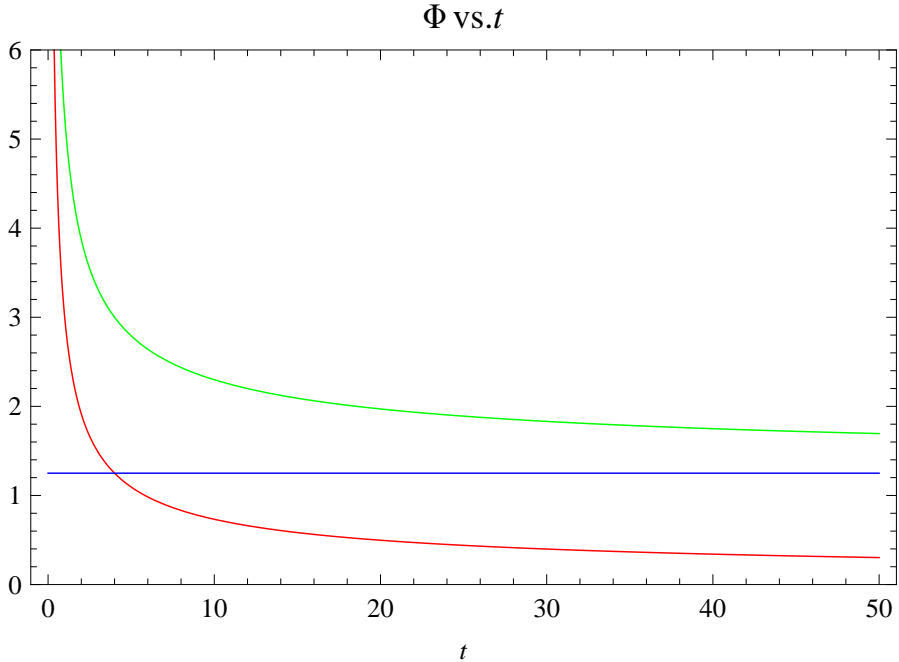}
\caption{ Plot of $\Phi(t)$ vs $t$ in the absence of visible brane matter 
 ($A = 2 $ and $C_1 = 2\sqrt{2}$ (red curve)); in the presence of visible brane matter ($A = 3$ and $C_1 = 2\sqrt{2}$(green curve)); non-zero asymptotic value
of $\Phi(t)$ when brane matter is present (horizontal blue line).}
\end{figure}
Now in the presence of matter the proper distance between the branes using eqn.{(\ref{6})} is found to be :
\begin{equation}
d(t) = \frac{l}{2} \ln \left[ 1+\frac{C_1^2}{8t} \pm \frac{C_1 A}{2 \sqrt{2t}} +
\frac{A^2-4}{4}\right ]
\end{equation}
As $t\rightarrow\infty$, $d(t)$ is always non-zero and tends to a constant value for all $A>2$. 
%In this case, the variation of $\Phi(t)$ vs $t$ can be seen from Figure 2.  
%\begin{figure}
%\centerline{\epsfxsize=4in\epsffile{C_plot_2.eps}}
%\caption{ Plot of $\Phi(t)$ vs $t$ in the presence of visible brane matter. Here we take $A = 3$ and $C_1 = 1$}
%\end{figure}

\noindent Hence the proper distance never vanishes and therefore no instability exists. Thus, the perfect fluid matter on the brane with equation of state $p=\frac{\rho}{3}$ stabilizes the distance between the branes. It is to be noted that such an equation of state corresponds to a perfect fluid comprising of relativistic particles.

\subsection{Spatially curved solutions (k=-1,+1)}
\noindent Let us now construct the FRW solution on the visible brane with
non-zero spatial curvature. 
The Einstein equations given by eqn.{(\ref{20})} and eqn.{(\ref{9})} lead to an interesting observation--for $p=\frac{\rho}{3}$ the known Friedmann solutions 
for $k=+1$ and $k=-1$ survive. The scalar field equation remains unchanged but 
the scalar field profile is obviously different due to the different functional
forms of $a(t)$ for $k=+1$ and $k=-1$.\\
When $k=+1$, the addition of eqn.{(\ref{20})} and eqn.{(\ref{9})} (with
$p=\frac{\rho}{3}$ yields,
\beq
\frac{\ddot{a}}{a}\,+\,\frac{\dot{a}^2}{a^2}\,+\,\frac{1}{a^2}\,=\,0
\eeq
which on solving gives,
\begin{equation}
a(t) = \sqrt{A_1^2 - (t-A_1)^2}   \label{21}
\end{equation}
This is the well-known Friedmann scale factor where the universe begins
at $t=0$ and there is a big-crunch at $t=2A_1$.
Similarly for $k=-1$, addition of eqn.{(\ref{20})} and eqn.{(\ref{9})} results
in,
\beq
\frac{\ddot{a}}{a}\,+\,\frac{\dot{a}^2}{a^2}\,-\,\frac{1}{a^2}\,=\,0
\eeq
and the scale factor becomes,
\begin{equation}
a(t) = \sqrt{((t+A_1)^2-A_1^2)}   \label{22}
\end{equation}
With appropriate time translation ($t+A_1\rightarrow t$) , 
the solution of the scale factor may, in
general be written as,
\beq
a(t) = \sqrt{t^2 + K}
\eeq
where, $K$ is a real integration constant. In our form of the solution, 
we have chosen $K = -A_1^2 < 0$ and $a(0)=0$.
Here, the universe is eternally expanding, though with deceleration.

\noindent If we now write the scalar field as $ 1 + \Phi (t) = e^{\frac{2d(t)}{l}}$, then using this and  eqn.{(\ref{3})}, we can express the proper distance $d(t)$ in terms of integral of the scale factor as given below,
\begin{equation}
e^{\frac{d(t)}{l}}= \pm \frac{C_1}{2} \int \frac{dt}{a^3(t)} + B_1   \label{10}
\end{equation}
where $B_1$ is a constant of integration. Thus, given the scale factor for any spatial curvature $k= 0,-1,1$, eqn.{(\ref{10})} is the most general expression that determines the proper distance between the two $3$-branes. In order to verify whether a given scale factor always admits a non-zero $d(t)$ we need to verify 
that L. H. S. of the eqn.{(\ref{10})} is never be equal to one.\\
Substituting the the solution of the scale factor for $k=-1$ i,e. eqn.{(\ref{22})} in eqn.{(\ref{10})}, we get,
\begin{equation}
\sqrt{1+\Phi(t)} = e^{\frac{d(t)}{l}} = \mp \frac{C_1}{2 A_1^2} \frac{A_1 + t}{\sqrt{t(2A_1+t)}} + B  \label{23}
\end{equation}
Similarly, for $k=+1$ using eqn.{(\ref{21})} in eqn.{(\ref{10})} we get,
\begin{equation}
\sqrt{1+\Phi(t)} = e^{\frac{d(t)}{l}} = \pm \frac{C_1}{2 A_1^2} \frac{-A_1 + t}{\sqrt{t(t - 2A_1)}} + D  \label{24}
\end{equation}
where $B$ and $D$ are integration constants.\\[1mm]
Let us now try to see that if $\Phi(t)$ can become zero for any $t$. This will be possible for some $t$, if square of the R.H.S. of eqn.{(\ref{23})} and eqn.{(\ref{24})} become equal to $1$.\\
For $k = -1$, setting the square of the R.H.S. of eqn.{(\ref{23})} equal to one, we obtain the following roots for $t$ :
\beq
t_{\pm}\,=\,-1 \mp \sqrt{\dis\frac{1}{1 - \frac{C_1 ^2}{4 B'}}}
\eeq
where we have set $A_1 =1 $ ( without any loss of generality) and $B' = (\pm 1 - B) \geq 0$. Thus, if $C_1 > 4 B'$, the roots are complex conjugates and hence $\Phi(t)$ is never zero. For $C_1^2 < 4 B'$, there is a positive root for which $\Phi(t)$ can become zero. However, choosing $B=1$ and the upper sign in $B'$ one may eliminate this possibility too.\\[1mm] 
Similarly, for $k=1$, we can obtain the roots for $t$ when $\Phi(t)$ may become zero. These turn out to be (with $A_1 =1 $ and $D' = (\pm 1 - D)^2$),
\beq
t_{\pm}\,=\,1 \mp \sqrt{\dis\frac{1}{1 + \frac{C_1 ^2}{4 D'}}}
\eeq
Here, it is clear that both roots lie within the domain of $t$. which is $0 \leq t \leq 2$. If $D' = 0$ (i,e. $D = 1$, with the upper sign in the expression for $D'$) then there ia a single root at $t = 1$.
The variation of radion field $\Phi(t)$ with time for both $k = -1, 1$ are 
shown in Figure 2 and they confirm the above discussion. It is clear that in the $k=+1$ case an instability
(brane collision) arises during the evolution of the universe.\\
The condition under which $d(t)$ can be never equal to one for the spatially 
flat case has already been shown earlier.

\begin{figure}[h]
\begin{minipage}{18pc}
\includegraphics[width=18pc]{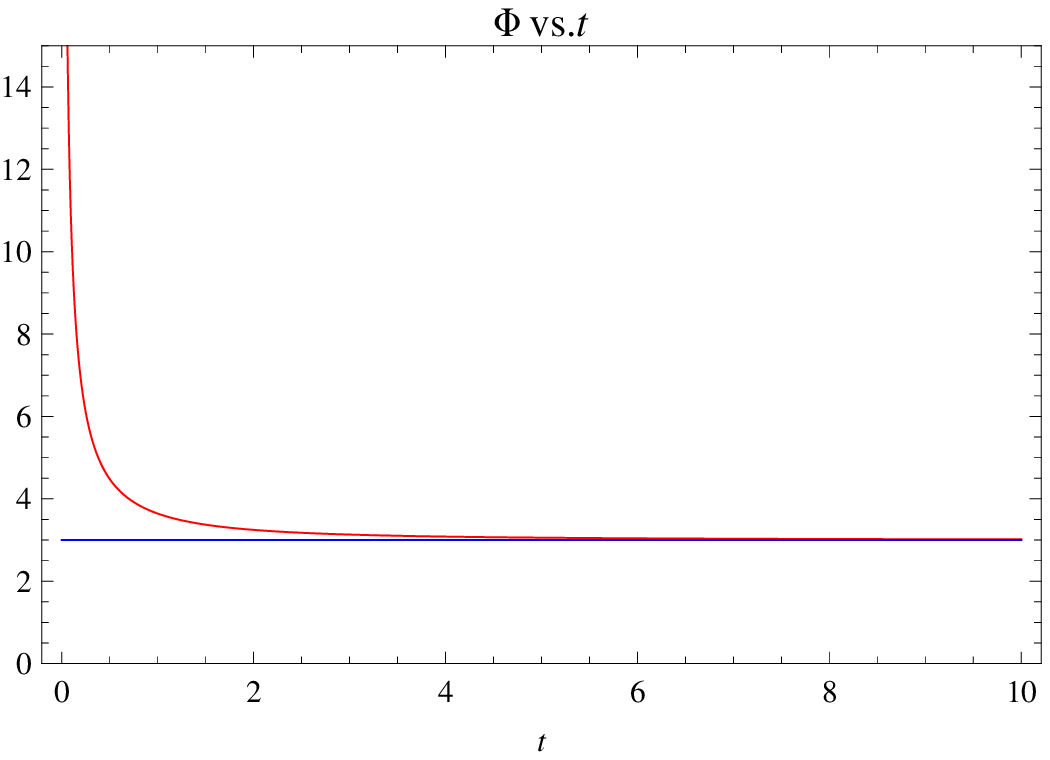}
\caption{$\Phi(t)$ vs. $t$ for $k=-1$; $C_1=2, A_1=1, B=1$. Blue line shows
asymptotic value.}
\end{minipage}\hspace{2pc}%
\begin{minipage}{18pc}
\includegraphics[width=18pc]{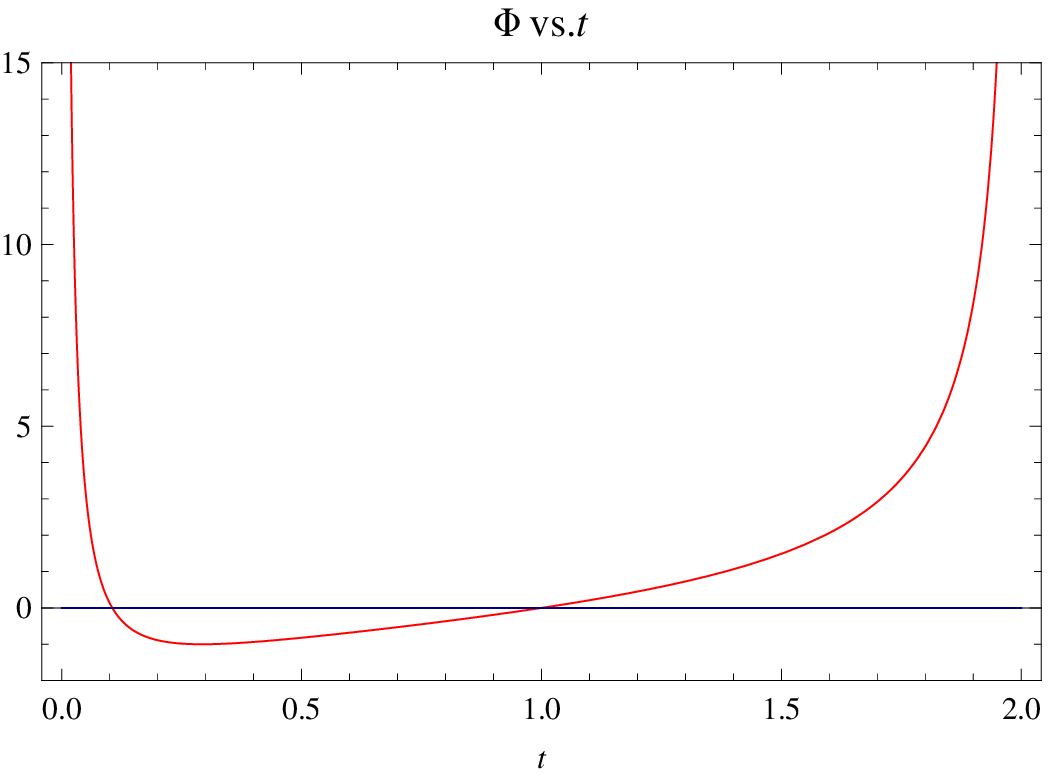}
\caption{$\Phi(t)$ vs. $t$ for $k=+1$; $C_1=2, A_1=1, D=1$. $\Phi(t)$
equals zero at $t=1$.}
\end{minipage}
\end{figure}

\section{ Spherically symmetric, static solutions}

\noindent 
Let us now look at spherically symmetric static solutions of the 
effective Einstein's equations on the visible brane. In constructing 
such a solution, it is legitimate to assume a radial co-ordinate i.e. $r$ 
dependent 
radion field $\Phi(r)$. We begin with a line element of the 
Majumdar--Papapetrou  \cite{mp} form which uses isotropic coordinates:
\begin{equation}
ds^2=-\frac{1}{U^2(r)} dt^2 +U^2(r) \left [ dr^2 + r^2 d\theta^2
+r^2\sin^2\theta d\phi^2 \right ]   \label{11}
\end{equation}
where $U(r)$ is the unknown function to be determined by solving Einstein's equations. First, let us assume that the branes are empty i,e. $T^{a}\,_{\mu \nu} = T^{b}\,_{\mu \nu} = 0$. Substituting the metric ansatz given by eqn.{(\ref{11})} in eqn.{(\ref{EE})}, we arrive at the following field equations :
\beq
-2 \frac{U''}{U} +\left (\frac{U'}{U}\right )^2 - 4\frac{U'}{Ur} =
-\frac{\Phi'^2}{4\Phi(1+\Phi)} + \frac{U'}{U}\frac{\Phi'}{\Phi}    \label{12}
\eeq
\beq
-\left (\frac{U'}{U}\right )^2 =  
-\frac{3\Phi'^2}{4\Phi(1+\Phi)} - \frac{U'}{U}\frac{\Phi'}{\Phi} -
\frac{2\Phi'}{\Phi r}     \label{13}
\eeq
\beq
\left (\frac{U'}{U}\right )^2 =  
\frac{\Phi'^2}{4\Phi(1+\Phi)} + \frac{U'}{U}\frac{\Phi'}{\Phi} +
\frac{\Phi'}{\Phi r}    \label{14}
\eeq
Here, a prime denotes a derivative with respect to $r$. 
Adding eqn.{(\ref{13})} and eqn.{(\ref{14})} one obtains,
\begin{equation}
\frac{\Phi'}{\Phi}\left ( \frac{\Phi'}{2(1+\Phi)} + \frac{1}{r} \right )\,=\,0   \label{15}  
\end{equation}
Since $\Phi'(r)\neq 0$ one can consider the term in brackets in the 
above equation as a condition on $\Phi$ and its derivative.
However, the scalar field equation for $\Phi(r)$ given by,
\begin{equation}
\Phi'' + 2\frac{\Phi'}{r} = \frac{\Phi'^2}{2(1+\Phi)}
\end{equation}
can be readily integrated once to get
\begin{equation}
\frac{\Phi'}{\sqrt{1+\Phi}} = \frac{2 \,C_1}{r^2}    \label{16}
\end{equation}
where $C_1$ is a positive, non-zero integration constant.
Consistency of eqn.{(\ref{15})} (i.e. the equation
$\frac{\Phi'}{2(1+\Phi)} + \frac{1}{r}=0$) and eqn.{(\ref{16})} 
for $\Phi'(r)$ leads to a unique form of $\Phi (r)$ given by :
\begin{equation}
\Phi (r) = \frac{C_1^2}{r^2}-1
\end{equation}
Further, we can use the condition in eqn.{(\ref{15})} to rewrite the 
Einstein equations in the following form:
\beq
-2 \frac{U''}{U} +\left (\frac{U'}{U}\right )^2 - 4\frac{U'}{Ur} =
\frac{\Phi'}{2\Phi r} + \frac{U'}{U}\frac{\Phi'}{\Phi} \label{17}
\eeq
\beq
-\left (\frac{U'}{U}\right )^2 =  
-\frac{\Phi'}{2\Phi r} - \frac{U'}{U}\frac{\Phi'}{\Phi}   \label{18}
\eeq
\beq
\left (\frac{U'}{U}\right )^2 =  
\frac{\Phi'}{2\Phi r} + \frac{U'}{U}\frac{\Phi'}{\Phi}   \label{19}
\eeq
We note that the R. H. S. of the above field equations lead to
the traceless-ness requirement on the L. H. S. Therefore $U(r)$ must satisfy the following differential equation:
\begin{equation}
U'' + 2 \frac{U'}{r} =0
\end{equation}
which is the Laplace equation $\nabla^2 U=0$ expressed in spherical polar coordinates (this result is the same as what follows in Einstein-Maxwell theory for Majumdar-Papapetrou type solutions \cite{mp}). The solution for $U(r)$ is therefore 
straightforward and is given by,
\begin{equation}
U(r) = C_2-\frac{C_3}{r}
\end{equation}
where $C_2$ and $C_3$ are two positive, non-zero constants. Substituting the solutions obtained for $U(r)$, $\Phi(r)$ and their derivatives
in either of the two Einstein's equations, i.e. eqn.{(\ref{17})} or 
eqn.{(\ref{18})}, we find a single condition between the non-zero constants 
given as :
\begin{equation}
C_1^2 C_2^2 = C_3^2   \label{vac_con}
\end{equation}
Hence the final solutions for $U(r)$ and $\Phi(r)$ in terms of $C_1, C_2$ and $C_3$ become :
\beq
\Phi (r) = \frac{C_3^2}{C_2^2 r^2} - 1
\eeq
\beq
U(r) = C_2-\frac{C_3}{r}
\eeq
At $r= C_1 = \dis\frac{C_3}{C_2}$, $U(r)=0$ which implies existence of a 
black hole horizon. Now for the same value of $r$, the radion field 
$\Phi(r)$ or the inter-brane distance vanishes suggesting an instability which needs to be removed. To keep $\Phi(r)$ always non-zero, we apply the method 
adopted in the case of cosmology (see the earlier section of this 
article).
We add traceless matter on the visible brane. Therefore, 
using eqn.{(\ref{11})} in eqn.{(\ref{EE})} once again (but with the
presence of matter on the visible brane) we now 
obtain the following Einstein equations on the visible brane,
\begin{eqnarray}
-2 \frac{U''}{U} +\left (\frac{U'}{U}\right )^2 - 4\frac{U'}{Ur} =
-\frac{\Phi'^2}{4\Phi(1+\Phi)} + \frac{U'}{U}\frac{\Phi'}{\Phi} +\frac{\kappa^2}{l\Phi} \rho \\
-\left (\frac{U'}{U}\right )^2 =  
-\frac{3\Phi'^2}{4\Phi(1+\Phi)} - \frac{U'}{U}\frac{\Phi'}{\Phi} -
\frac{2\Phi'}{\Phi r} +\frac{\kappa^2}{l\Phi} \tau \\
\left (\frac{U'}{U}\right )^2 =  
\frac{\Phi'^2}{4\Phi(1+\Phi)} + \frac{U'}{U}\frac{\Phi'}{\Phi} +
\frac{\Phi'}{\Phi r} +\frac{\kappa^2}{l\Phi} p
\end{eqnarray}
where $\rho (r)$, $\tau(r)$ and $p(r)$ are the diagonal components (in the
frame basis)
of the energy--momentum tensor on the visible brane.  
As long as this additional brane matter is traceless, i.e.
\begin{equation}
-\rho+\tau+2 p =0
\end{equation}
there is no change in the scalar field differential equation. 
The general solution of the scalar field equation however needs to be
taken as
\begin{equation}
\Phi(r) = \left (\frac{C_1}{r} +\frac{C_4}{2}\right )^2 -1  \label{25}
\end{equation}
where $C_4$ is a positive constant which is responsible for 
generating the brane matter. 
Even though with $C_4=0$ the $r$-dependent $\Phi (r)$ produces a 
non-flat on-brane metric, it involves an unstable radion and 
also corresponds to the case when the visible brane is empty. 
We can easily see that as long as $C_4>2$, $\Phi(r)$ never vanishes 
and by having traceless matter on the visible brane, the 
instability disappears for this particular, spherically symmetric solution 
with a $r$-dependent inter-brane distance $\Phi(r)$.
\begin{figure}
\centerline{\epsfxsize=6in\epsffile{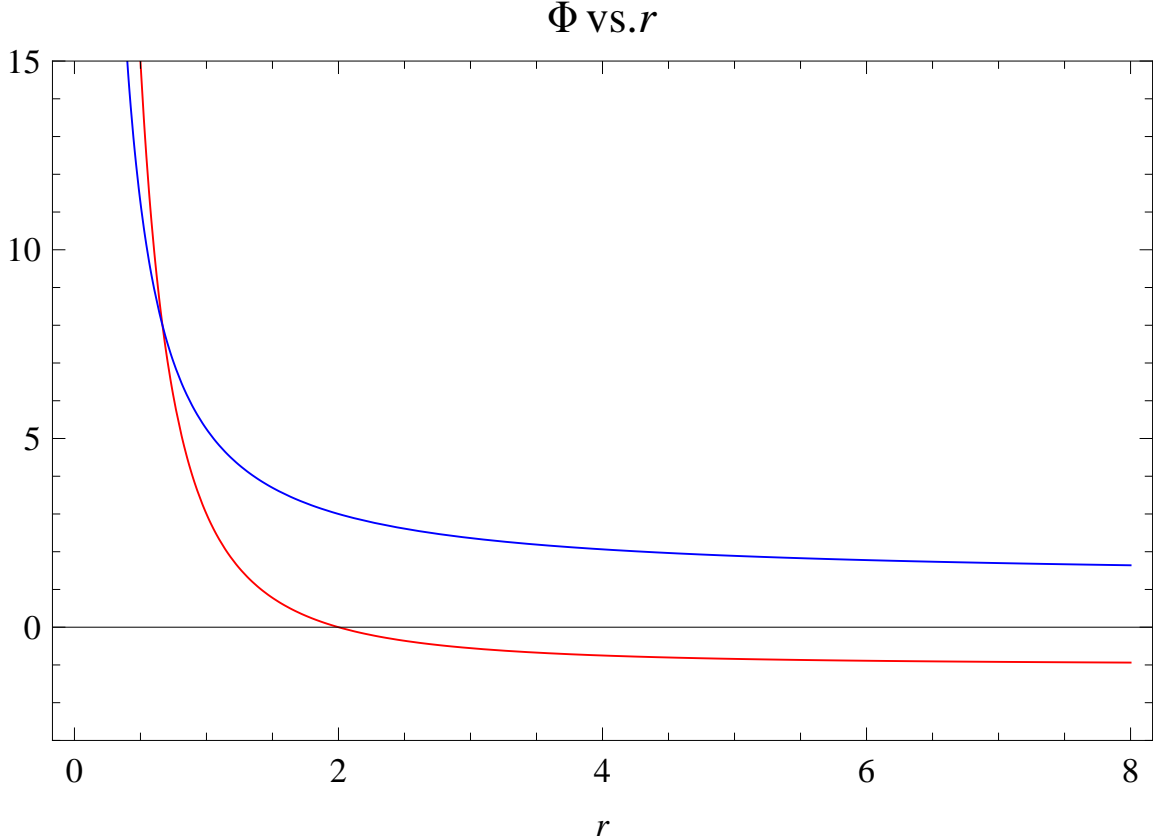}}
\caption{ Plot of $\Phi(r)$ vs $r$ when $C_1 = 2, C_4=0$ (red curve) 
and $C_1=2, C_4=3$ (blue curve)}
\end{figure}
It is to be noted further that the solution for $U(r)$ remains unaltered 
under the tracelessness condition on brane matter. However, it is now possible
to choose $\frac{C_3}{C_2}$ to be different from $C_1$. We assume
\begin{equation}
U(r) = 1-\frac{C_5}{r}
\end{equation}
From the above expressions for $U(r)$
and $\Phi(r)$, the visible brane matter energy momentum, i.e. $\rho$, $\tau$ 
and $p$ turn out to be:
\begin{eqnarray}
\rho = \frac{l}{\kappa^2} \frac{1}{\left (1-\frac{C_5}{r}\right )^2}\frac{1 }{r^4} \left (C_5 C_1 C_4 + C_1^2 - C_5^2
+\frac{C_5^2 C_4^2}{4} \right ) \\
\tau = \frac{l}{\kappa^2} \frac{1}{\left (1-\frac{C_5}{r}\right )^2}\left (\frac{3 C_1 C_4 C_5 - C_1^2+ C_5^2-\frac{C_5^2 C_4^2}{4} }{r^4} -\frac{2 C_1 C_4}{r^3} - \frac{2 C_1 C_4 C_5^2}{r^5} \right )\\
p = \frac{l}{\kappa^2}  \frac{1}{\left (1-\frac{C_5}{r}\right )^2} \left (\frac{- C_1 C_4 C_5 + C_1^2- C_5^2+\frac{C_5^2 C_4^2}{4}}{r^4} + \frac{C_1 C_4}{r^3}
+\frac{C_1 C_4 C_5^2}{r^5}\right )
\end{eqnarray}
We note
that $C_5$ as well as $C_1$ cannot be zero in order to ensure a non-constant $U(r)$ and $\Phi(r)$. At the same time, $C_4 = 0$ is
also not desirable because it would lead to an instability (i.e. $\Phi(r)$ becoming zero at some r).
Further, all the three constants must satisfy 
$C_1 > 0$, $C_4 > 0$ and $C_5 > 0$. It is possible to have 
both $C_1$  and $C_4$ negative but this does not effect the functional forms of $\rho$, $\tau$ and $p$ or $\Phi(r)$. However,  if
one chooses $C_5 < 0$, the solution leads to a naked singularity. 
It is also clear that we cannot have $\tau  = p $ because this condition
leads to a quadratic equation for $r$ which implies specific $r$ values as
its solutions. The only allowed
condition is the one for traceless matter, i.e. $\rho = \tau + 2p$.
In addition, the Weak Energy Condition (WEC) or Null Energy Condition (NEC) 
will be violated . In particular,
\begin{equation}
\rho + \tau = - \frac{l}{\kappa^2}\frac{2 C_1 C_4}{r^4}
\end{equation}
Since we must have $C_1,C_4 > 0$  for stability, $\rho + \tau < 0$ but
one can satisfy $\rho>0$ and $\rho+p>0$ by choosing the constants 
appropriately. Even though the $\rho$, $\tau$ and $p$ violate WEC and NEC, 
the `effective  matter' which is equal to the {\em total} expressions 
in the R. H. S. of Eqns. (51)-(53)
does satisfy the WEC, NEC. One can easily check this by
renaming the quantities on the R. H. S. of (51)-(53) as $\rho_{eff}$, $\tau_{eff}$, $p_{eff}$ and verifying the validity of  
$\rho_{eff}>0$, $\rho_{eff}+\tau_{eff}=0$ and $\rho_{eff}+p_{eff}>0$.
 The functional forms of $\rho$, $\tau$ and $p$ are shown in Figure 5 for
a specific choice of the parameters, with $C_1=C_5$. We have also checked
(not shown here) that the profiles of $\rho$, $tau$ and $p$ are similar
when $C_1\neq C_5$.
\begin{figure}
\centerline{\epsfxsize=6in\epsffile{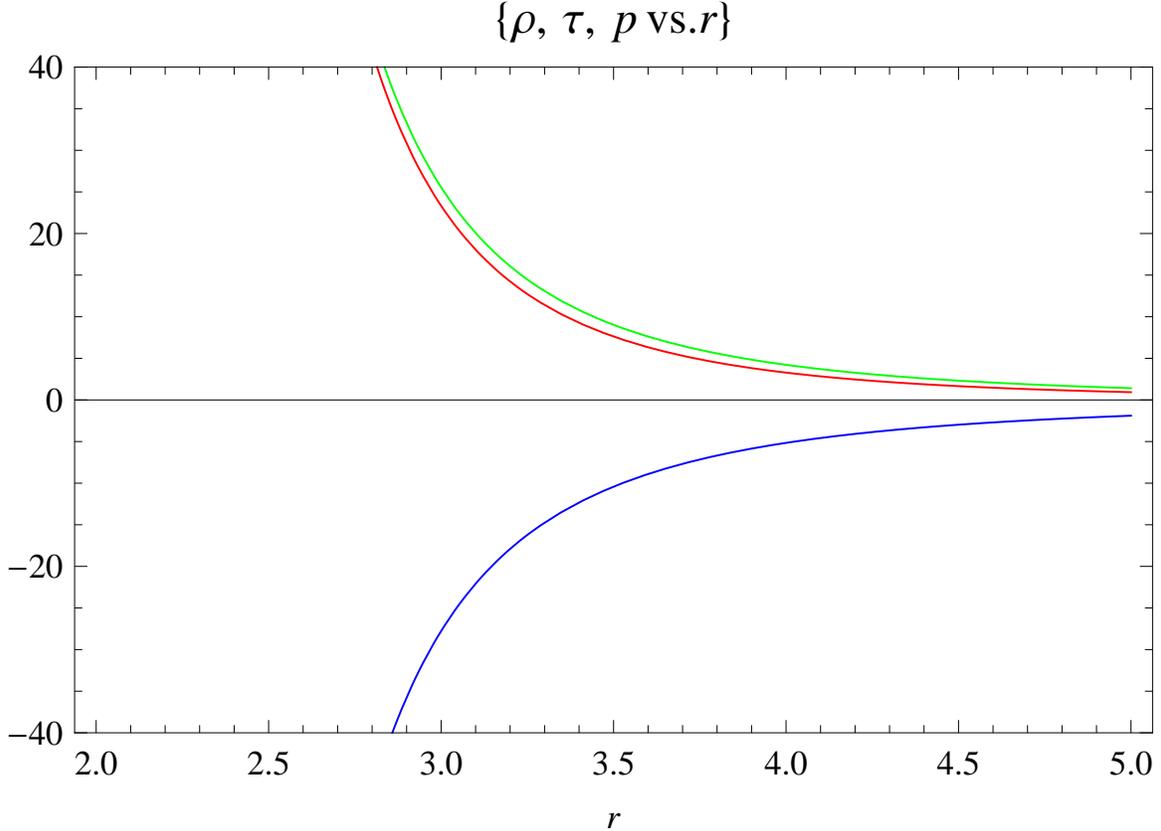}}
\caption{ Plot of $\frac{\kappa^2}{3l} \rho(r)$, $\frac{\kappa^2}{3l} \tau(r)$, $\frac{\kappa^2}{3l} p(r)$ vs $r$ in red, blue and green, respectively. Here
$C_1 = 2 $ and $C_4 = 3$ and the $y$ axis values are scaled by factor of 10.}
\end{figure}
It is now easy to convert the metric solution (and the scalar field solution) 
into the extremal Reissner--Nordstrom black hole form by the 
following identifications:
\begin{equation}
r=r'-M \hspace{0.2in};\hspace{0.2in} C_1=M
\end{equation}
This leads to the extremal 
Reissner--Nordstrom black hole metric given as:
\begin{equation}
ds^2 = -\left (1-\frac{M}{r'}\right )^2 dt^2 +\frac{dr'^2}{\left (1-\frac{M}{r'}\right )^2} + r'^2 \left (
d\theta^2 +\sin^2\theta d\phi^2\right )
\end{equation}
We note that $r'=M$ is the location of horizon as well as the spacetime
singularity.

\noindent For such spherically symmetric solutions, 
we can also obtain the $\Psi(r)$ by exploiting the relation
between $\Phi(r)$ and $\Psi(r)$ given in \cite{kanno}. For example,
in the simple case (without visible brane matter) we have
\begin{eqnarray}
\Psi(r) = 1-\frac{r^2}{C_1^2}\\
h_{ij} = \frac{C_1^2}{r^2} f_{ij}
\end{eqnarray}
where the $h_{ij}$ is the metric on the Planck brane and  
the visible brane metric functions, $f_{ij}$, are given in terms of the 
$U(r)$ obtained above.  

\section{Conclusion}

\noindent In summary, we have shown the following:

$\bullet$ In the cosmological case, for traceless matter ($p=\frac{\rho}{3}$) on the visible brane
we find analytic solutions for the scale factor and the radion field. 
In the spatially flat universe, the scale factor is that of the radiation
dominated FRW case while the radion is stable and
never zero. Instability arises when there is no on-brane matter.
In a spatially curved universe with traceless, radiative matter, the
results are similar for the case of 
negative spatial curvature. With positive spatial curvature, instabilities
arise even with on-brane matter.  

$\bullet$ In the spherically symmetric, static case, in isotropic coordinates,
we find that the solution obtained is nothing but the extremal Reissner--Norstrom solution. However, there is no physical charge or mass here (like in
Einstein-Maxwell theory) and the radion field
parameters play the role of an equivalent charge or mass.

\noindent For the case when the matter on the brane is not necessarily 
traceless we are unable to find analytical solutions. 
Numerical work (not discussed here) suggest that the nature of the
solutions for say, $p=0$ or $p=-\rho$ are 
different from the  solutions for $p=\frac{\rho}{3}$ discussed here.

\noindent It is noteworthy that our analytic solutions are all
obtained using traceless, on-brane matter. However, we also note
that the stability of the radion may not necessarily have any connection
with the tracelessness of on-brane matter, though the need for some
on-brane matter to achieve stability has been demonstrated in our
examples. A hint about what kind of matter can achieve stability of the
radion can be obtained by setting $C_4=0$ in the expressions for $\rho$, $\tau$
and $p$. Notice (from Eqns. (57)-(59)) that for $C_1^2>C_5^2$ the
NEC and WEC will be satisfied. Does this indicate that a stable radion
requires energy-condition violating on-brane matter? 
A general statement is unlikely here though one may surely try to explore 
the exact link between the nature of
on-brane matter and radion stability in future investigations.

\noindent Finallly, the fact that we have rediscovered known solutions 
(i.e. the FRW scale factors in cosmology and the extremal Reissner--Nordstrom
in the static, spherisymmetric case)
in the
context of a theory different from General Relativity is certainly
welcome. This feature was also noticed in the first analytic solution
in the Shiromizu-Maeda-Sasaki on-brane, effective theory \cite{sms} where the
Reissner--Nordstrom solution was rediscovered as an exact solution \cite{rn}.
There, the interpretation of a charge or mass was entirely geometric
and largely dependent on the presence of the extra dimensions. Here too,
it is the presence of extra-dimensions, through the space or time
dependent radion, which is responsible for the nature of the solutions, though
on-brane matter seems to be crucial is maintaining stability.

\end{document}